\input psfig
\documentstyle[twoside,fleqn,espcrc2]{article}

\newcommand{\AmS}{{\protect\the\textfont2
  A\kern-.1667em\lower.5ex\hbox{M}\kern-.125emS}}

\title{Modelling the Quark Determinant in Full  QCD Simulations}
\author{ A.~Duncan\address{Dept. of Physics and Astronomy,University of Pittsburgh,
 Pittsburgh, PA 15260},%
  E.~Eichten\address{ Fermilab, PO Box 500, Batavia, IL60510},%
  and 
  H.~Thacker\address{Dept. of Physics, University of Virginia, 
 Charlottesville, VA 22901}}
\begin{document}
\begin{abstract}
The computational requirements and dynamics of Monte Carlo simulations
 of unquenched QCD incorporating the infrared quark eigenmodes (up to 
  $\approx\Lambda_{QCD}$) exactly and UV modes via a loop representation are
 discussed. The accuracy of such a loop representation is studied
 for a variety of lattice volumes and quark masses. The method has
 been successfully applied for lattices up to 10$^3$x20 at $a\simeq$0.17F
 with improved (clover) action, and allows simulations at or near 
 kappa critical.
 \end{abstract}

\maketitle

\section{Splitting the Quark Determinant}

  The essence of the truncated determinant approach \cite{truncdet} to unquenched
 QCD lies in the realization that the infrared part of the quark determinant (specifically,
 the determinant $\cal{D}(A)$ of the hermitian operator $H\equiv\gamma_{5}(D\!\!/(A)-m)$) can
 be gauge-invariantly split off, leaving an ultraviolet part which is accurately fit
 by a linear combination of a small number of Wilson loops. The eigenvalues 
$\lambda_{i}$ of $H$\\
(1) measure quark off-shellness (for $A=0$, $\lambda_{i}\rightarrow\pm\sqrt{p^{2}+m^{2}}$),\\
(2) are gauge-invariant, $\lambda_{i}(A)=\lambda_{i}(A^{g})$.

Thus we can write $\cal{D}(A)=\cal{D}_{IR}(A)\cal{D}_{UV}(A)$, where the infrared
 part $\cal{D}_{IR}(A)$ is defined as the product of the lowest $N_{\lambda}$ positive
 and negative eigenvalues of $H$, with $|\lambda_{i}|\leq\Lambda_{QCD}$
 (typically, $\simeq$ 300-400 MeV). This cutoff is chosen (a) to include as much as
 possible of the important low-energy chiral physics of the unquenched theory while
 (b) leaving the fluctuations of $\ln \cal{D}_{IR}$ of order unity after each sweep updating
 all links with the pure gauge action. This ensures that the acceptance rate is sufficiently high
 when the infrared determinant only is used in the accept/reject stage of the procedure.
 The crucial point is that it is possible to achieve {\em both} (a) and (b) on fairly large
 lattices (up to physical volume $\simeq$ 20 F$^4$) as well as at kappa values 
 arbitrarily close to kappa critical.

\section{Efficient Computation of the IR spectrum in QCD4}

 The following Lanczos procedure allows us to extract the needed infrared eigenvalues of
 $H$ relatively rapidly:\\
(1) Starting from an initial vector $v_1$, an orthonormal sequence $v_1,v_2,v_3,..v_{N_L}$ is generated by the standard recursion:\\
$\;\;\;\;v_{k+1}=\frac{1}{\beta_{k}}Hv_k-\frac{\beta_{k-1}}{\beta_{k}}v_{k-1}-\frac{\alpha_{k}}{\beta_{k}}v_{k}$\\
with the constants $\alpha_{k},\beta_{k}$ determined from overlaps of generated vectors. In the
 basis of the $v_{i}$, $H$ is tridiagonal. The corresponding real symmetric
 tridiagonal matrix $T_{N_L}$ has the $\alpha_{k}$ on the diagonal and the $\beta_{k}$ on the 
 sub (and super) diagonal. \\
(2) A Cullum-Willoughby sieve \cite{Cullum} is used to identify and remove spurious eigenvalues.\\
(3)  The remaining ``good" eigenvalues converge most rapidly in the least dense part of
 the spectrum, in particular, in the needed infrared portion. This remains true even at 
 kappa critical ( in marked contrast to the {\em matrix inversions} needed in HMC simulations,
 for example), where the density of eigenvalues near zero is still small. 
 The stability and accuracy of the converged
 eigenvalues has been checked extensively by gauge transforming the gauge field. \\
(4) The diagonalization of the $T_{N_L}$ matrix (typically, of order 10,000 in the QCD case)
 can be completely parallelized using the Sturm sequence property \cite{golub}
 of tridiagonal matrices in which  nonoverlapping parts of the spectrum are independently    extracted by a bisection procedure.

\section{Fitting the UV modes- Loop representations of the Quark determinant}

 The effective gauge action generated by internal quark loops is a gauge-invariant 
 functional, which can be evaluated explicitly in a hopping parameter expansion:
\begin{eqnarray*}
  \ln {\cal D}(A)/V &=&  288\kappa^{4}\sum L_{1} +2304\kappa^{6}\sum L_{2}\\
 &+&4608\kappa^{6}\sum L_{3} + 1536\kappa^{6}\sum L_{4}+..
\end{eqnarray*}
 where $V$ is the lattice volume, $L_1$ a generic plaquette, and $L_{2,3,4}$ are
 6 link loops with link directions $(i,i,j,-i,-i,-j)$, $(i,j,k,-j,-i,-k)$, and
$(i,j,k,-i,-j,-k)$ ($1\leq i,j,k \leq 4$).
 Unfortunately, this hopping parameter expansion is useless \cite{looppaper}
 except for extremely heavy
 quarks ($\kappa\rightarrow 0$), due to the contribution of large loops. Instead, large loops
 may be cut off by removing the IR modes $|\lambda_{i}|<\Lambda_{QCD}$. Now the
 expansion converges much more quickly, and we may write $\ln {\cal D}_{UV}(A) = V\sum L_{i}(A)$
 where the sum involves only a small number of loops (as we shall see, typically less than 10)
 and the coefficients $c_{i}$ are determined nonperturbatively. Of course, to compute 
 ${\cal D}_{UV}$, we need the complete spectra for an ensemble of configurations. The
 computational cost for extracting a complete Dirac spectrum via Lanczos is large but manageable, as
 this calculation need only be done for a limited number of decorrelated configurations.
 For example, the complete Dirac spectrum for a 10$^3$x20 lattice has 240,000 eigenvalues
 and requires about 500,000 Lanczos sweeps, equivalent to about 1 400-Mhz-Pentium-week.
 The final spectrum can be checked with analytic spectral sum rules which give the sum
 of powers of the eigenvalues as explicit functionals of small loops. 

  The results of a fit of  $\ln {\cal D}_{UV}$ to a linear combination of Wilson loops on 
 an ensemble of 10$^3$x20 lattices at beta=5.7 (with clover improvement) and $\kappa$=0.1425
 is shown in Fig.1  The configurations were generated including the truncated determinant
 $\ln {\cal D}_{IR}$ and therefore already contain the exact low-energy chiral physics. The
 fit is very good once 4, 6 and 3 figure-8 8-link operators are included in the fit . These results
 suggest that the full determinant can be accurately modelled by computing the low eigenvalues
 exactly and including the remaining high modes via an approximate loop action. In any event,
 we expect that the UV fluctuations affect primarily the scale of the theory, while for the low
 energy spectrum, quark off-shellness is limited to about $\Lambda_{QCD}$ and dimensionless
 mass ratios should therefore be largely insensitive to ${\cal D}_{UV}$. 
\begin{figure}
\psfig{figure=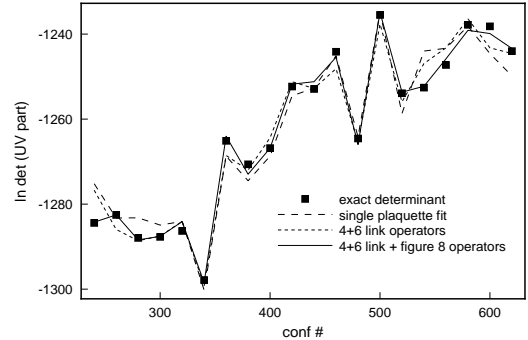,
width=0.95\hsize}
\vspace{-0.35in}
\caption{Fit of UV Determinant to Loop Action, $\kappa$=0.1425}
\label{fig:detfit}
\end{figure}

\section{MonteCarlo Dynamics for QCD4 Simulations with the Truncated Determinant}

   In the truncated determinant approach to QCD4 \cite{truncdet}, unquenched configurations
 are generated by the following algorithm:\\
\begin{enumerate}
\item Update the gauge configuration with the pure gauge action, using a procedure
 compliant with detailed balance. We have used Metropolis link updates applied to 
 a randomly chosen block of noninterfering links to ensure detailed balance while
 maintaining parallelizability of the computation. 
\item Apply a metropolis accept/reject criterion based on the the effective quark
 action
\begin{equation}
   S_{\rm quark} = N_{F}\ln{\rm det} {\cal D}_{IR}(A)\;\;\;(N_{F}=2)
\end{equation}
\end{enumerate}
 Typically, we  use an IR cutoff for the truncated determinant corresponding to 
 a gauge-invariant eigenvalue of 300-400 MeV.
 This procedure leads to tolerable acceptance rates on lattices of  fairly large physical
 volume - we have explored systems up to 20 $F^4$, while controllable finite size errors
 in electromagnetic fine structure lattice studies \cite{empaper} with a long range massless
 U(1) field require lattice volumes $\geq$6 $F^4$. The simulations in progress are on
 three different lattice sizes:\\
(1) 10$^3$x20 lattices at $\beta$=5.7 (clover improved) at $\kappa$=0.1415, 0.1425, 0.1436
 and 0.1440 (the last value being very close to kappa critical). This corresponds to a
 physical volume of roughly (1.7F)$^3$x3.4F.\\
(2) $6^4$ lattices at an effective $\beta$ of 4.5, but using the O(a$^2$) improved gauge action of
 Alford et al,\cite{impgauge}, at kappa critical. \\
(3)  $8^4$ lattices at an effective $\beta$=4.5 (O(a$^2$) improved), again at kappa critical. 
   
  For the 10$^3$x20 lattices we have checked that the truncated determinant simulations succeed in equilibrating
 the configurations, and that they decorrelate reasonably rapidly subsequent to 
 equilibration.  The equilibration can be studied by looking at the relaxation of
 $S_{\rm quark}$ from the quenched value corresponding to the starting configuration.
 Even at $\kappa=\kappa_{c}$, the 10$^3$x20 lattices equilibrate after a few hundred sweeps
 (see Fig. 2). 

\begin{figure}
\psfig{figure=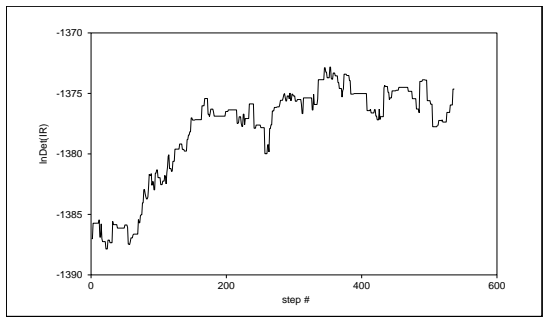,
width=0.95\hsize}
\vspace{-0.35in}
\caption{Determinant relaxation, $\kappa$=0.1440}
\label{fig:detrelax}
\end{figure}

 To measure decorrelation we have calculated the autocorrelation of the pion 
 propagator at various time separations, keeping configurations separated
 by 20 steps of the basic algorithm. A typical example,  from the 10$^3$x20
 runs with $\kappa$=0.1425, is shown in Fig.3, where the pion correlator at time
 slice 6 is seen to be effectively decorrelated after about 30 steps of the algorithm.
\begin{figure}
\psfig{figure=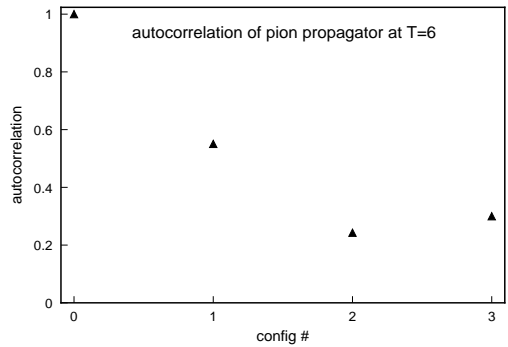,
width=0.95\hsize}
\vspace{-0.35in}
\caption{Decorrelation of pion propagator, $\kappa$=0.1425}
\label{fig:piondecorr}
\end{figure}

   Cases (2,3) represent physically large lattices (33 F$^4$ and 105 F$^4$ resp.) at 
 kappa critical and display critical slowing down (several thousand sweeps
 are needed to equilibrate the configurations). However the lattice sizes are
 small: new configurations can be generated and the truncated determinant
 computed quite rapidly ($\simeq$ 20 minutes for the 6$^4$ lattices, two hours
 for the 8$^4$ lattices, on a Pentium 400 Mhz processor). These lattices will allow
 a detailed study of string-breaking and other unquenched dynamical effects (see
 talk of Eichten, this conference).

\vspace{-0.1in}


\end{document}